\documentclass[a4paper,12pt]{article}
\usepackage{amsfonts}
\usepackage{latexsym}
\usepackage{amsmath}
\usepackage{amssymb}
\usepackage{amssymb}
\usepackage{slashed}
\usepackage{upgreek}
\usepackage{mathrsfs}
\usepackage{bbm}

\usepackage{relsize}
\usepackage{graphicx}

\newcommand{\newsection}{    
\setcounter{equation}{0}\section}
\def\appendix#1{\addtocounter{section}{1}\setcounter{equation}{0}
\renewcommand{\thesection}{\Alph{section}}
\section*{Appendix \thesection\protect\indent \parbox[t]{11.15cm}{#1}}
\addcontentsline{toc}{section}{Appendix \thesection\ \ \ #1}}

\def\bbe{{\bf{e}}}
\font\mybb=msbm10 at 11pt

\def\bb#1{\hbox{\mybb#1}}

\def\bZ {\bb{Z}}
\def\bR {\bb{R}}

\def\bC {\bb{C}}

\def\gX{\Gamma\mkern-4.0mu X}
\def\gom{\Gamma\mkern-4.0mu \omega}

\def\sX{\slashed {X}}
\def\sgX{\slashed {\gX}}

\newcommand{\bea}{\begin{eqnarray}}
\newcommand{\eea}{\end{eqnarray}}

\begin{document}
\begin{titlepage}
\begin{center}
\vspace*{-1.0cm}
\hfill DMUS--MP--17/04 \\

\vspace{2.0cm} {\Large \bf On supersymmetric AdS$_6$ solutions in 10 and 11 dimensions} \\[.2cm]

\vskip 2cm
  J. B.  Gutowski$^1$ and G. Papadopoulos$^2$
\\
\vskip .6cm

\begin{small}
$^1$\textit{Department of Mathematics,
University of Surrey \\
Guildford, GU2 7XH, UK \\
Email: j.gutowski@surrey.ac.uk}
\end{small}\\*[.6cm]

\begin{small}
$^2$\textit{  Department of Mathematics, King's College London
\\
Strand, London WC2R 2LS, UK.\\
E-mail: george.papadopoulos@kcl.ac.uk}
\end{small}\\*[.6cm]

\end{center}

\vskip 3.5 cm
\begin{abstract}

We prove  a non-existence theorem for smooth, supersymmetric,  warped   $AdS_6$ solutions with  connected, compact without boundary  internal space in
  $D=11$ and  (massive) IIA  supergravities. In IIB supergravity we show that if such $AdS_6$ solutions exist,  then the NSNS and RR 3-form fluxes must
   be linearly independent and certain spinor bi-linears must be appropriately restricted. Moreover  we demonstrate that the internal space admits an   $\mathfrak{so}(3)$ action   which  leaves all the fields invariant and for smooth solutions the   principal orbits  must have co-dimension two. We also describe the topology and geometry of internal spaces that admit such a $\mathfrak{so}(3)$ action and show that there are no solutions for which the internal space has topology $F\times S^2$, where $F$ is an oriented surface.

\end{abstract}

\end{titlepage}


\section{Introduction}

AdS spaces have found widespread applications first as compactifications of supergravity theories and more recently as a tool
to explore superconformal field theories in the context of the AdS/CFT correspondence, for reviews see \cite{duff, maldacena}. As AdS/CFT  provides a correspondence between $AdS_n$ backgrounds of
10 and 11-dimensional supergravity theories with conformal field theories in $(n-1)$-dimensions, properties of superconformal theories can be
investigated  in the context of supergravity theories. This has given a new impetus to understanding the AdS backgrounds that preserve a fraction
of the spacetime supersymmetry.

In this context, and in particular for $AdS_6$ backgrounds, several  solutions have been  found in \cite{andy, lozano1, lozano2, tomasiello, Kim, dhoker, dhoker2} and explored in the context
of AdS$_6$/CFT$_5$, see eg \cite{seiberg, seiberg2, aharony, bergman, jafferis}.  Furthermore it has been shown in \cite{mads, iibads, iiaads} that $AdS_6$ backgrounds preserve either 16 or 32 supersymmetries
in all 10- and 11-dimensional supergravity theories.  Moreover it is known for sometime that these theories do not admit maximally supersymmetric  backgrounds \cite{maxsusy} which are strictly
locally isometric to $AdS_6$.
Some additional  non-existence results have been established
in \cite{tomasiello, passias} for smooth $AdS_6$ solutions preserving 16 supersymmetries in 11-dimensional and IIA supergravities under  the assumption that the Killing spinors factorize into Killing spinors of AdS and Killing spinors on the internal space and
some additional restrictions on the internal spaces.

It has been demonstrated in \cite{nfact} that the Killing spinors of AdS backgrounds do not factorize into Killing spinors of AdS and Killing spinors on the internal space, and so requiring factorization
 is an additional assumption. Related to this some care is required in establishing no-go theorems for AdS backgrounds as  $AdS_n$ spaces can be written as warped products of $AdS_k$, $k<n$
 and so $AdS_n$ backgrounds can be re-interpreted as warped $AdS_k$ solutions \cite{strominger}-\cite{slicex}. For example the $AdS_7$ maximally supersymmetric background of 11-dimensional supergravity
 can be re-interpreted as a warped maximally supersymmetric $AdS_6$ solution. To exclude such a scenario, we put some global assumptions on the internal spaces of $AdS_6$ backgrounds
 that we shall describe below.

In this paper, we shall prove a non-existence theorem for smooth warped $AdS_6$ backgrounds in 11-dimensional and  (massive) IIA  imposing only as
assumptions\footnote{In particular, we do not assume that the Killing spinors are factorized as described above.} that  the internal space is closed\footnote{For simplicity, we shall refer to smooth AdS backgrounds with closed internal space as ``smooth closed AdS backgrounds''.}, ie it is compact and without boundary.  As this theorem for maximally supersymmetric backgrounds has already been demonstrated,
the main focus is to establish the result for $AdS_6$ solutions preserving 16 supersymmetries.

Furthermore we shall demonstrate  some non-existence results for smooth closed $AdS_6$ backgrounds in IIB supergravity provided some additional assumptions are made. In turn these assumptions
can be viewed as necessary conditions for the existence of IIB $AdS_6$ solutions.
In particular, we demonstrate that if smooth  closed $AdS_6$ IIB backgrounds exist, the NSNS and RR 3-form fluxes must be linearly independent.  This rules out the existence
of smooth closed solutions with only NSNS or RR 3-form fluxes and all their $SL(2,\bR)$ duals. We also find that this linear independence condition on the 3-form fluxes is met
provided some spinor bilinears are appropriately restricted. In particular, a certain (twisted) scalar bilinear must vanish, another (twisted) scalar bilinear must be somewhere vanishing on the
 internal space  and a 2-form bilinear must have rank at most two. The full set of conditions is summarized in section 4.4.  Another necessary condition is that closed $AdS_6$ IIB backgrounds
 must always have active scalars.

 Next we show that the internal spaces of all $AdS_6$ IIB backgrounds admit
a non-trivial\footnote{This means  that the associated Killing vector fields do not vanish everywhere on the internal space.} $\mathfrak{so}(3)$ action with leaves all the fields invariant.  If the infinitesimal $\mathfrak{so}(3)$ action can be integrated to an effective $SU(2)$ or $SO(3)$ action and   $AdS_6$ IIB backgrounds are smooth and closed, then the principal orbits must be of co-dimension 2.  This rules out
all solutions  for which $\mathfrak{so}(3)$ acts on the internal space with co-dimension 1 principal orbits.  The diffeomorphic type of internal manifolds in the oriented case can be specified by utilizing the classification results of \cite{parker}.  These include the  spin manifolds
\bea
S^4~,~~~  p (S^1\times S^3)\# q (S^1\times \bR P^3)~,~~~M(F)~,
\label{diffma}
\eea
where $M(F)$ is the unique spin oriented 2-sphere
bundle over a surface $F$, see also appendix C.  In particular if $F$ is oriented, then $M(F)=F\times S^2$. Next we demonstrate with a partial integration argument that there are
   no smooth closed $AdS_6$ solutions that have internal spaces with topology $F\times S^2$.

The methodology that will be followed has been developed in \cite{mads, iibads, iiaads} for  investigating all $AdS$ backgrounds.  In particular, it is assumed that the metric and fluxes
are invariant under the isometries of the $AdS$ space but otherwise there are no additional assumptions like an ansatz on the form of the Killing spinors. Then the supergravity KSEs are integrated  along the AdS space which gives the dependence of the Killing spinor
on the AdS coordinates as well as a set of KSEs along the internal space.  These   are associated with the gravitino and other algebraic KSEs of the original
 supergravity theory, and in addition  there is an  algebraic Killing spinor equation which arises as an integrability condition associated with the solution of the KSEs along the AdS.
The existence of smooth closed solutions is then explored by applying  techniques,  like that of the Hopf maximum principle, and taking into account the assumptions made
on the topology of the internal space. It turns out that this methodology is sufficient to prove the non-existence  of supersymmetric $AdS_6$ backgrounds  in  11-dimensional and (massive) IIA supergravities.  However in IIB, we have not been able to establish such a result. Instead, we have given some necessary conditions for the existence of smooth closed
 $AdS_6$ solutions, and specified their diffeomorphic type using  classification results
of \cite{parker} for 4-dimensional manifolds admitting an effective  $\mathfrak{so}(3)$ action.

This paper has been organized as follows.  In sections two and three, we establish the non-existence of smooth closed $AdS_6$ backgrounds for 11-dimensional and (massive) IIA  supergravities, respectively.
In section four, we investigate the existence of smooth closed $AdS_6$ solutions in  IIB  supergravity.
 In appendix A, we investigate the isometries of the internal space and state our conventions. In appendix B, we present some Fierz identities that have been used
 in our derivations. In appendix C, we summarize the results of \cite{parker} on the structure of 4-manifolds admitting a non-abelian group action and in appendix D, we
 present various formulae for the spinor bilinears of IIB $AdS_6$ backgrounds.

\newsection{ Warped $AdS_6$  backgrounds in D=11}

We begin by briefly summarizing the general structure of warped $AdS_6$ solutions in 11-dimensional supergravity, as determined
in \cite{mads}, whose conventions we shall follow throughout this section.
 The metric and 4-form which are invariant under the isometries of $AdS_6$ are given by
\begin{eqnarray}
ds^2 &=& 2 du (dr + r h )+ A^2 (dz^2+ e^{2z/\ell} \sum_{a=1}^3 (dx^a)^2) +ds^2(M^5)~,
\cr
F &=&   X~,
\label{mhmadsn}
\end{eqnarray}
where $ds^2(M^5)$ is the metric on the internal space $M^5$, and we have written the solution as a near-horizon geometry \cite{adshor},
with
\bea
h = -{2 \over \ell}dz -2 A^{-1} dA~.
\eea
The coordinates $(u,r, z, x^1, x^2, x^3)$ are those of the $AdS_6$ space,  $A$ is the warp factor which  is a function on  $M^5$,
 and $X$ is a closed 4-form on $M^5$. The metric $ds^2(M)$,
$A$ and $X$ depend only on the coordinates of $M^5$, $\ell$ is the radius of $AdS_6$.

The 11-dimensional Einstein equation implies that the warp factor satisfies the equation
\bea
\label{lapein}
D^k \partial_k \log A=-{5\over \ell^2} A^{-2}- 6 \partial^k \log A\, \partial_k \log A+{1\over 144}  X^2~,
\eea
where $D$ is the Levi-Civita connection on $M^5$.
The remaining components of the Einstein and gauge field equations
are listed in \cite{mads}, however we shall only require ({\ref{lapein}})
for the analysis that follows.
In particular, ({\ref{lapein}}) implies that $A$ is everywhere non-vanishing
on $M^5$, on assuming that $M^5$ is connected and all fields are smooth.

\subsection{The Killing spinors}

The KSEs of 11-dimensional supergravity can be solved along $AdS_6$ yielding
\begin{eqnarray}
\epsilon&=&\sigma_+-\ell^{-1} \sum_{a=1}^{3} x^a\Gamma_{az} \tau_++ e^{-{z\over\ell}} \tau_++\sigma_-+e^{{z\over\ell}}(\tau_--\ell^{-1} \sum_{a=1}^{3} x^a \Gamma_{az} \sigma_-)
\cr
&&-\ell^{-1} u A^{-1} \Gamma_{+z} \sigma_--\ell^{-1} r A^{-1}e^{-{z\over\ell}} \Gamma_{-z}\tau_+~,
\label{kkk}
\end{eqnarray}
where the spinors $\sigma_\pm$ and $\tau_\pm$ are Majorana $Spin(10,1)$ spinors that depend only on the coordinates $y^I$ of $M^5$ and satisfy the light-cone   projections
\begin{eqnarray}
\Gamma_\pm \sigma_\pm =0~,~~~\Gamma_\pm\tau_\pm=0~.
\label{dec}
\end{eqnarray}
The gamma matrices have been adapted to the  spacetime frame
\begin{eqnarray}
\label{orthfr}
&&\bbe^+ = du~, \qquad \bbe^- = dr+rh~, \qquad
\bbe^z=A dz~, \qquad \bbe^a = A e^{z/\ell} dx^a~,~~~
\cr
&&
\bbe^i= e^i_I dy^I~,~~~
\eea
where $ds^2(M^5)=\delta_{ij} e^i_I e^j_J dy^I dy^J$.
The expression for the Killing spinor $\epsilon$ (\ref{kkk}) is derived after intergrating the KSEs along $AdS_6$. In particular notice that
we do not assume that the Killing spinor $\epsilon$ factorizes as a Killing spinor  on $AdS_6$ and a Killing spinor on $M^5$ which is an additional
assumption on the form of the Killing spinors \cite{mads, nfact}.

The remaining independent Killing spinor equations (KSEs) are
\bea
D^{(\pm)}_i \sigma_\pm=0~,~~~D^{(\pm)}_i \tau_\pm=0~,
\label{kseadsk1}
\eea
and
\bea
{\Xi}^{(\pm)} \sigma_\pm=0~,~~~{\Xi}^{(\mp)} \tau_\pm=0~,
\label{kseadsk2}
\eea
where
\bea
D^{(\pm)}_i&=&D_i \pm {1\over 2} \partial_i \log A-{1\over 288} \sgX_i
+{1\over 36} \sX_i~,
\cr
\Xi^{(\pm)}&=&-{1\over2}\Gamma_z \Gamma^i \partial_i \log A\mp {1\over 2\ell}  A^{-1} +{1\over 288}\Gamma_z \sX~.
\eea
The (\ref{kseadsk1}) KSEs are a suitable restriction of the gravitino KSE of 11-dimensional supergravity on $M^5$ while  the (\ref{kseadsk2}) conditions
arise during the integration process of the KSEs along $AdS_6$.
Notice that the   algebraic KSEs ({\ref{kseadsk2}}) imply that $\sigma_+$ and $\tau_+$ cannot be linearly dependent. In fact, we shall later demonstrate that they
must be orthogonal. For our Clifford algebra conventions see \cite{mads}.

\subsection{Counting the Killing Spinors} \label{count1}

The counting of supersymmetries of warped $AdS_6$ backgrounds will be given in 11-dimensional supergravity. A similar counting applies
to (massive) IIA and IIB supergravities and so it
 will not be repeated  below for these theories.

To begin, note that if
$\sigma_+$ is a solution of the $\sigma_+$ KSEs, then so is $\Gamma_{ab} \sigma_+$ for $a,b=1,2,3$.
Furthermore, $\tau_+=\Gamma_{za}  \sigma_+$
are also solutions to the $\tau_+$  KSEs.  The eight spinors $\sigma_+, \Gamma_{ab} \sigma_+, \Gamma_{za}  \sigma_+, \Gamma_{123z} \sigma_+$ are linearly independent.

The spinors
$\sigma_-$, $\tau_-$ can also be constructed from $\sigma_+$ and $\tau_+$. This is
because if  $\sigma_+, \tau_+$ is a solution, then so is
$\sigma_-=A \Gamma_{-z}  \sigma_+, \tau_-=A \Gamma_{-z}  \tau_+$
and  conversely, if $\sigma_-, \tau_-$ is a solution, then so is
$\sigma_+=A^{-1} \Gamma_{+z} \sigma_-, \tau_+=A^{-1} \Gamma_{+z} \tau_-$.  Thus all Killing spinors of $AdS_6$
backgrounds are generated by the $\sigma_+$ Killing spinors.

As a result the number of Killing spinors of $AdS_6$ backgrounds is a multiple of 16. Thus if there are $AdS_6$ solutions of supergravity
theories, they will either preserve 16 or 32 supersymmetries.  It has been shown  sometime ago that 11-dimensional, (massive) IIA and IIB supergravities
do not admit maximally supersymmetric $AdS_6$ solutions \cite{maxsusy}. As a result, it remains to investigate the $AdS_6$ backgrounds that preserve
16 supersymmetries.

\subsection{ Proof of the main theorem  in $D=11$ supergravity}

\subsubsection{Orthogonality of $\tau_+$ and $\sigma_+$ spinors}

Before we proceed with the proof of the main theorem, we shall first establish the orthogonality of $\tau_+$ and $\sigma_+$ Killing spinors.
It will be convenient to define
\bea
W = {\tilde{\star}} X~,
\eea
where ${\tilde{\star}}$ denotes the Hodge dual on $M^5$, so $W$ is a 1-form.
To proceed, we set
\bea
\Lambda = \sigma_+ + \tau_+~,
\eea
and using  ({\ref{kseadsk1}}), we find
\bea
\label{max11a}
D_+ \parallel \Lambda \parallel^2 = -A^{-1} D_i A \parallel \Lambda \parallel^2 +{1 \over 6} W_i \langle \Lambda, \Gamma_{(4)} \Lambda \rangle~,
\eea
where $\Gamma_{(4)}$ denotes the product of the gamma matrices
in the 4 directions of $AdS_6$ spanned by ${\bf{e}}^z$ and
${\bf{e}}^{\rm a}$, with the convention that
\bea
\Gamma_{(4)} \Gamma_{i_1 i_2 i_3 i_4 i_5}
\phi_\pm = \pm \epsilon_{i_1 i_2 i_3 i_4 i_5} \phi_\pm~.
\eea
Using the algebraic KSE ({\ref{kseadsk2}}), ({\ref{max11a}}) can be rewritten as
\bea
\label{max11b}
D_i \parallel \Lambda \parallel^2 = -{2 \over \ell} A^{-1} \langle
\sigma_+, \Gamma_i \Gamma_z \tau_+ \rangle
\eea
and the gravitino KSE ({\ref{kseadsk1}}) implies that
\bea
\label{max11c}
D^i \langle \sigma_+, \Gamma_i \Gamma_z \tau_+ \rangle =
- \langle \sigma_+, \Gamma_i \Gamma_z A^{-1} D^i A \tau_+ \rangle~.
\eea
On taking the divergence of ({\ref{max11b}}), and utilizing
({\ref{max11c}}), we find
\bea
\label{max11d}
D^i D_i \parallel \Lambda \parallel^2 +2 A^{-1} D^i A D_i \parallel \Lambda \parallel^2 =0~.
\eea
A maximum principle argument then implies that
\bea
\parallel \Lambda \parallel^2={\rm const}~.
\label{Lcon}
\eea
Note that for the application of the maximum principle it is sufficient to assume that the backgrounds are smooth and $M^5$ is closed.
As $\parallel \Lambda \parallel^2$ is constant,  ({\ref{max11a}}) and ({\ref{max11b}}) imply
\bea
\label{max11aux1}
-A^{-1} D_i A \parallel \Lambda \parallel^2 +{1 \over 6} W_i \langle \Lambda, \Gamma_{(4)} \Lambda \rangle =0~,
\eea
and
\bea
\label{max11aux2}
\langle
\sigma_+, \Gamma_i \Gamma_z \tau_+ \rangle=0~,
\eea
respectively.

Next taking inner products of the algebraic KSE ({\ref{kseadsk2}})
acting on $\sigma_+$ and $\tau_+$ with $\tau_+$ and $\sigma_+$ respectively,
one finds, on using ({\ref{max11aux2}}), that
\bea
-{1 \over 2\ell} \langle \tau_+, \sigma_+ \rangle -{1 \over 12}
\langle \tau_+, \slashed{W} \Gamma_z \Gamma_{(4)} \sigma_+ \rangle &=&0~,
\nonumber \\
{1 \over 2\ell} \langle \sigma_+, \tau_+ \rangle -{1 \over 12}
\langle \sigma_+, \slashed{W} \Gamma_z \Gamma_{(4)} \tau_+ \rangle &=&0~.
\eea
Subtracting these expressions, we deduce that
\bea
\label{d11orthog}
\langle \sigma_+, \tau_+ \rangle =0~.
\eea
This establishes the orthogonality of $\sigma_+$ and $\tau_+$ Killing spinors.

\subsubsection{A non-existence theorem} \label{11nex}

Next, note that as a consequence of how the $\sigma_+$, $\tau_+$ spinors
are generated from each other as described in the previous subsection,
it follows that $\Gamma_{(4)} \sigma_+ = \tau_+'$, and hence
({\ref{d11orthog}}) implies that
\bea
\label{d11van}
\langle \sigma_+, \Gamma_{(4)} \sigma_+ \rangle =0~.
\eea
So, on substituting $\Lambda=\sigma_+$ into ({\ref{max11aux1}}),
we find
\bea
dA=0~,
\eea
and so the warp factor is constant.
The gravitino KSE ({\ref{kseadsk1}}) also implies that
\bea
D_i \langle \sigma_+ , \Gamma_{(4)} \sigma_+ \rangle = {1 \over 6} W_i \parallel \sigma_+ \parallel^2~,
\eea
and hence this expression together with ({\ref{d11van}}) also give that
\bea
W=0~,
\eea
and hence $X=0$. So, we have proven that for $AdS_6$ solutions,
one has
\bea
dA=0, \qquad X=0~.
\eea
However, as  $A$ is constant and $X=0$, the field equation ({\ref{lapein}}) for the warp factor $A$  does not admit a solution.  This proves the theorem.

To summarize, combining the result\footnote{For the classification of maximally supersymmetric solutions in \cite{maxsusy}
there is no need to impose smoothness and compactness conditions. The proof works in general and the backgrounds are classified
up to a local isometry.} of \cite{maxsusy} with the proof described above, one concludes that {\it there are no smooth closed
warped $AdS_6$ solutions of 11-dimensional supergravity}.

One question that arises is whether some of the assumptions we have made can be lifted. First our result can be generalized somewhat.  For this notice that
we need the smoothness assumption as well as the restrictions on the internal space $M^5$ to demonstrate that the only solution of  (\ref{max11d}) is that
 $\Lambda$ is of constant length (\ref{Lcon}).
Thus we can replace all these assumptions with (\ref{Lcon}).

The theorem is not valid if one removes all conditions on the fields and the internal space. This is because  locally $AdS_{n+1}$ spaces can be written as
warped products of $AdS_n$ spaces\footnote{This was  established in \cite{strominger} for $AdS_2$ and $AdS_3$ spaces explored further in \cite{emparan}-\cite{slicex}  and used
in the context of supersymmetric  $AdS$ backgrounds  in \cite{nfact}.}.  As a result  the $AdS_7\times S^4$ maximally supersymmetric
solution of 11-dimensional supergravity can be written locally as $AdS_6\times_w (\bR\times S^4)$ and so it can be interpreted as an $AdS_6$ solution.
This demonstrates that 11-dimensional supergravity admits maximally supersymmetric $AdS_6$ solutions.  However such solutions are not a contradiction to our theorem
as their internal space is not compact, see also \cite{nfact}.

\newsection{Warped ${AdS}_6$  backgrounds in  IIA supergravity }

The non-vanishing  fields of (massive)  IIA supergravity  for  warped ${AdS}_6 \times_w {M}^{4}$ backgrounds in the conventions of \cite{iiaads} are
\bea
ds^2 &=& 2 \bbe^+\bbe^-+ A^2 \big(dz^2+ e^{2z/\ell} \sum_{a=1}^{3}(dx^a)^2\big) +ds^2(M^{4})~,
\cr
G &= &G~,~~~
H=H~,~~~F=F~,~~~\Phi=\Phi~,~~~S=S~,
\eea
where  $A$ is the warp factor, $\Phi$ is the dilaton, $S$ is related to the cosmological constant,   $F$ and  $G$ are the 2-form and 4-form R-R field strengths correspondingly,
and  $H$ is a 3-form the NS-NS 3-form field strength. $A$, $\Phi$ and $S$ are functions on the  internal space $M^4$,
while $F$, $G$ and $H$ are 2-, 4- and 3-forms on $M^4$; all of them  depend only on the coordinates of $M^4$ as well as the metric of the internal space $ds^2(M^{4})$.
We have also introduced the frame $(\bbe^+, \bbe^-, \bbe^z, \bbe^a, \bbe^i)$ as in (\ref{orthfr})
 and $\ell$ is the radius of $\mathrm{AdS}_6$.
It will be convenient to define
\bea
H_{ijk} = \epsilon_{ijk}{}^\ell W_\ell, \qquad G_{ijk\ell}=X \epsilon_{ijk\ell}
\eea
where $W$ is now a 1-form on $M^4$ and $X$ is a function on $M^4$.

 The Bianchi identities of the (massive) IIA supergravity give
\bea
\label{iiabian}
 \nabla^i W_i = 0~,~~~
 dS = S d\Phi~,~~~
 dF = d\Phi \wedge F + S \star W~,
\eea
where $\star$ denotes the Hodge dual on $M^4$.
Furthermore, the field equations of (massive) IIA supergravity give
\begin{align}
\label{iiafeq}
 \nabla^2 \Phi &= -6 A^{-1} \partial^i A \partial_i \Phi + 2 ( d\Phi )^2 + \frac{5}{4} S^2 + \frac{3}{8} F^2 - {1 \over 2} W^2 + {1 \over 4}X^2~,
 \nonumber \\
 dW &= -6 A^{-1} dA \wedge W +2 d \Phi \wedge W + S \star F + X F~,
 \nonumber \\
 \nabla^j F_{i j} &= -6 A^{-1} \partial^j A F_{i j} + \partial^j \Phi F_{i j} +X W_i~,
 \nonumber \\
 dX &= X \big(-6 A^{-1} dA + d \Phi \big)~,
\end{align}
and the Einstein equation separates into an AdS component
\begin{equation}
\label{warpeq}
 \nabla^2 \ln A = -5 \ell^{-2} A^{-2} - 6 A^{-2} ( dA )^2 + 2 A^{-1} \partial_i A \partial^i \Phi +{1 \over 4} X^2 + \frac{1}{4} S^2 + \frac{1}{8} F^2 ,
\end{equation}
which is interpreted as the field equation for the warp factor. The Bianchi identity $dS=S d \Phi$ implies that if $S$ is smooth, then either $S$
is nowhere vanishing on $M^4$ or $S \equiv 0$ everywhere on $M^4$.  In what follows, we shall use the conventions and methodology of \cite{iiaads}
for the investigation of $AdS$ spaces where more details can be found.
%
%

\subsection{The Killing Spinors}

The solution of the KSEs of (massive) IIA supergravity along $AdS_6$ can be expressed as in (\ref{kkk}),
where $\sigma_\pm$ and $\tau_\pm$ depend only on the coordinates of $M^4$ and are Majorana $Spin(9,1)$ spinors that satisfy the lightcone
projections $\Gamma_\pm\sigma_\pm=\Gamma_\pm \tau_\pm=0$.  The remaining KSEs have been stated in \cite{iiaads}.
For the analysis which follows, it suffices to consider those acting on
$\chi= \sigma_+$ and $\chi=\tau_+$.
The IIA gravitino KSE implies that
\bea
\label{iiagrav}
\nabla_i \chi = \bigg(-{1 \over 2} A^{-1} \partial_i A
+{1 \over 4} \Gamma_{(4)} \slashed{\Gamma W}_i -{1 \over 8}S \Gamma_i
-{1 \over 16} \slashed{F} \Gamma_i \Gamma_{11}
-{1 \over 8}X {\tilde{\Gamma}}_{(4)} \Gamma_i \bigg) \chi~,
\eea
where
\bea
\Gamma_{(4)}= \Gamma_{zxyw}, \qquad \Gamma_{ijk\ell}=\epsilon_{ijk\ell}
{\tilde{\Gamma}}_{(4)} \ .
\eea
Furthermore the IIA  dilatino KSE implies that
\bea
\label{iiaalg1}
\bigg(\slashed{\partial}\Phi -{1 \over 2} \slashed{W} \Gamma_{(4)}
+{5 \over 4} S +{3 \over 8} \slashed{F}\Gamma_{11}
+{1 \over 4}X {\tilde{\Gamma}}_{(4)} \bigg) \chi =0
\eea
and there is a further algebraic KSE which arises during the integration of the KSEs along $AdS_6$ given by
\bea
\label{iiaalg2}
\bigg(-{c \over 2 \ell}A^{-1}\Gamma_z -{1 \over 2} A^{-1} \slashed{\partial}A -{1 \over 8}S -{1 \over 16}\slashed{F} \Gamma_{11}
-{1 \over 8} X {\tilde{\Gamma}}_{(4)} \bigg) \chi=0~,
\eea
where $c=1$ for $\chi=\sigma_+$ and $c=-1$ for $\chi=\tau_+$.  This is the analogue of the (\ref{kseadsk2}) in $D=11$ supergravity.
Note that (\ref{iiaalg2}) implies that $\sigma_+$ and $\tau_+$ are linearly independent.

The counting of supersymmetries of IIA $AdS_6$ backgrounds proceeds as in the $D=11$ case and so IIA $AdS_6$ backgrounds preserve
either 16 or 32 supersymmetries.  As there are not maximally supersymmetric $AdS_6$ backgrounds in IIA supergravity \cite{maxsusy},
it remains to investigate the $AdS_6$ backgrounds preserving 16 supersymmetries.

\subsection{Proof of the main theorem}

\subsubsection{Orthogonality of $\sigma_+$ and $\tau_+$ spinors}

As in $D=11$ supergravity, we proceed by setting
\bea
\Lambda = \sigma_+ + \tau_+~,
\eea
then ({\ref{iiagrav}}), together with the algebraic conditions
({\ref{iiaalg2}}) imply that
\bea
\label{iiamax1}
\nabla_i \parallel \Lambda \parallel^2 ={2 \over \ell} A^{-1} \langle
\tau_+, \Gamma_i \Gamma_z \sigma_+ \rangle~.
\eea
Furthermore,  ({\ref{iiagrav}}) also implies that
\bea
\label{iiamax2}
\nabla^i \langle \tau_+, \Gamma_i \Gamma_z \sigma_+ \rangle
=- \langle \tau_+, \Gamma_i \Gamma_z A^{-1} \nabla^i A \sigma_+ \rangle~.
\eea
On taking the divergence of ({\ref{iiamax1}}) and using ({\ref{iiamax2}}),
we find
\bea
\label{iiamax3}
\nabla^2 \parallel \Lambda \parallel^2 +2 A^{-1} \nabla^i A \nabla_i
\parallel \Lambda \parallel^2 =0~.
\eea
An application of the maximum principle implies that
\bea
\parallel \Lambda \parallel^2 = {\rm const}~.
\eea
For this, it is sufficient to require that  the solutions are smooth and
the internal space $M^4$ is closed.

Hence ({\ref{iiamax1}}) implies that
\bea
\label{iiamax4}
\langle
\tau_+, \Gamma_i \Gamma_z \sigma_+ \rangle=0~.
\eea
Returning to the algebraic conditions ({\ref{iiaalg2}}); on taking inner products and making use of ({\ref{iiamax4}}),  we obtain
\bea
-{1 \over 2 \ell} A^{-1} \langle \tau_+, \sigma_+ \rangle
-{1 \over 8}S \langle \tau_+, \Gamma_z \sigma_+ \rangle
-{1 \over 16} \langle \tau_+, \Gamma_z {\slashed{F}} \Gamma_{11}
\sigma_+ \rangle -{1 \over 8}X \langle \tau_+, \Gamma_z
{\tilde{\Gamma}}_{(4)} \sigma_+ \rangle &=&0
\nonumber \\
{1 \over 2 \ell} A^{-1} \langle \sigma_+, \tau_+ \rangle
-{1 \over 8}S \langle \sigma_+, \Gamma_z \tau_+ \rangle
-{1 \over 16} \langle \sigma_+, \Gamma_z {\slashed{F}} \Gamma_{11}
\tau_+ \rangle -{1 \over 8}X \langle \sigma_+, \Gamma_z
{\tilde{\Gamma}}_{(4)} \tau_+ \rangle &=&0
\nonumber \\
\eea
On subtracting these expressions, one obtains
\bea
\label{iiaorthog}
\langle \sigma_+, \tau_+ \rangle=0~.
\eea
This establishes the orthogonality of $\sigma_+$ and $\tau_+$ spinors.

\subsubsection{Additional properties of KSEs}

To establish some additional properties of the KSEs observe that if $\sigma_+$ is a Killing spinor, then $\Gamma_{(4)} \sigma_+$ solves the KSEs as a $\tau_+$
Killing spinor.  This follows from the relation between $\sigma_+$ and $\tau_+$ Killing spinors as explained in the context of 11-dimensional supergravity theory,
 section \ref{count1}, that also applies  in IIA supergravity.  Then it follows from the orthogonality condition ({\ref{iiaorthog}}) of $\sigma_+$ and $\tau_+$ spinors that
\bea
\label{iiavan}
\langle \sigma_+, \Gamma_{(4)} \sigma_+ \rangle =0~.
\eea
To proceed further, on eliminating the $F$ terms between the algebraic KSEs
({\ref{iiaalg1}}) and ({\ref{iiaalg2}}), and using ({\ref{iiavan}}), one obtains the condition
\bea
\label{algax1}
\bigg({1 \over 6}\partial_i \Phi - {1 \over 2} A^{-1}
\partial_i A \bigg) \parallel \sigma_+ \parallel^2
+{1 \over 12} S \langle \sigma_+, \Gamma_i \sigma_+ \rangle =0~.
\eea
The KSEs also imply that
\bea
\label{scala2}
\nabla_i \langle \sigma_+, \Gamma_{(4)} \sigma_+ \rangle
= \bigg({1 \over 3} \partial_i \Phi - A^{-1} \partial_i A \bigg) \langle
\sigma_+, \Gamma_{(4)} \sigma_+ \rangle
-{1 \over 6} W_i \parallel \sigma_+ \parallel^2 +{1 \over 6} S
\langle \sigma_+, \Gamma_{(4)} \Gamma_i \sigma_+ \rangle~,
\nonumber \\
\eea
which together with ({\ref{iiavan}}) gives
\bea
\label{algax2}
W_i \parallel \sigma_+ \parallel^2 = S
\langle \sigma_+, \Gamma_{(4)} \Gamma_i \sigma_+ \rangle~.
\eea

There are then two cases to consider depending on whether  $S \not\equiv 0$
or $S \equiv 0$.

\subsubsection{A non-existence theorem for standard IIA}

In the special case for which $S$ vanishes,  $S \equiv 0$, ({\ref{algax1}}) gives that
\bea
{1 \over 6} d \Phi - {1 \over 2} A^{-1} dA=0
\eea
and ({\ref{algax2}}) implies that
\bea
W=0~.
\eea
The dilaton field equation ({\ref{iiafeq}}) then becomes
\bea
\nabla^2 \Phi = {3 \over 8} F^2 + {1 \over 4} X^2~.
\eea
On integrating this expression over $M^4$, one finds
\bea
F=0, \qquad X=0
\eea
and also
\bea
d \Phi=0, \qquad dA=0~,
\eea
where again we have used that $M^4$ is closed.
However, the warp factor equation ({\ref{warpeq}}) then admits no solution which establishes the non-existence theorem.

\subsubsection{A non-existence theorem for massive IIA}

As we have already mentioned a consequence of the Bianchi identity $dS= S d\Phi$  is that $S$ is nowhere vanishing. Furthermore, left-multiplying  ({\ref{iiaalg1}}) with ${\tilde{\Gamma}}_{(4)}$, taking $\chi=\sigma_+$, and then
taking the inner product with $\sigma_+$,   implies that
\bea
\label{xs1}
S \langle \sigma_+, {\tilde{\Gamma}}_{(4)} \sigma_+ \rangle = - {1 \over 5} X \parallel \sigma_+ \parallel^2 \ .
\eea
Also, taking the inner product of ({\ref{iiaalg1}}), with $\chi=\sigma_+$,
and using ({\ref{algax1}}), ({\ref{algax2}}) and ({\ref{xs1}}) to eliminate
the spinor bilinear terms, one finds
\bea
\label{xs2}
-2(d \Phi)^2+6 A^{-1} \partial_i A \partial^i \Phi
-{1 \over 2} W^2+{5 \over 4} S^2 -{1 \over 20}X^2 =0~.
\eea
Then, using ({\ref{xs2}}) to eliminate the $S^2$ term, the dilaton
field equation ({\ref{iiafeq}}) can be rewritten as
\bea
\label{xs3}
\nabla^i \bigg(A^{12} \nabla_i \big( e^{-4 \Phi}\big) \bigg) = -4 A^{12} e^{-4 \Phi}
\bigg({3 \over 8}F^2 + {3 \over 10} X^2 \bigg) \ .
\eea
Integrating this expression over $M^4$ and using the assumption that $M^4$ is closed yields the conditions
\bea
\label{st1}
F=0, \qquad X=0~,
\eea
and ({\ref{xs3}}) then simplifies to
\bea
\nabla^2 \big(e^{-4 \Phi} \big) +12 A^{-1} \partial^i A \partial_i \big(e^{-4 \Phi}\big)=0~.
\eea
The maximum principle then implies that
\bea
\label{st2}
d \Phi =0 \ .
\eea
The algebraic KSE ({\ref{iiaalg2}}) then gives that
\bea
{1 \over 8}S \sigma_+ = \bigg(-{1 \over 2 \ell} A^{-1} \Gamma_z -{1 \over 2} A^{-1}
\slashed{\partial}A \bigg) \sigma_+~,
\eea
and on squaring this expression we find
\bea
\label{st3}
S^2= 16 \ell^{-2} A^{-2} + 16 A^{-2} (dA)^2 \ .
\eea
On substituting ({\ref{st1}}), ({\ref{st2}}) and ({\ref{st3}}) into the
Einstein equation ({\ref{warpeq}}), this condition can be rewritten as
\bea
\nabla^2 A^2 = -2 \ell^{-2} \ .
\eea
However, this equation admits no regular solution. Hence there are no smooth closed supersymmetric warped
$AdS_6$ solutions  in massive IIA supergravity.

The assumptions mentioned in the description of the non-existence theorem are essential. First observe that the maximum principle has been used to establish that
the length of the Killing spinor $\Lambda$ is constant as in the 11-dimensional case.  In addition, further partial integration
arguments are required to establish the result, which require topological restrictions on the internal space.
These assumptions can possibly  be weakened, but not entirely removed. This is because one can  reduce the $AdS_7\times S^4$ solution
of 11-dimensional supergravity, which has been interpreted as an $AdS_6$ solution in section (\ref{11nex}), along  Killing directions of $S^4$ to find   $AdS_6$
solutions in IIA supergravity. However the existence of such  solutions will not be a contradiction, as they do not satisfy the conditions of our theorem.

\newsection{ $AdS_6 \times _w M^4$ solutions in IIB supergravity}

The non-vanishing form fluxes  of IIB  $AdS_6$ backgrounds  have support on the internal space $M^4$.  In particular as the 5-form R-R field strength $F$ is self-dual, it vanishes.
The rest of the fields can be written   as
\begin{eqnarray}
ds^2= 2 \bbe^+ \bbe^-+ A^2( dz^2+e^{2z\over\ell} \sum_{a=1}^3 (dx^a)^2)+ ds^2(M^4)~,~~~G=H~,~~P=\xi~,
\end{eqnarray}
where $G$ is a twisted\footnote{In this formulation, the $G$ is twisted with respect to the pull-back of the $U(1)$ bundle associated with the upper-half plane $SU(1,1)/U(1)$ which is the target space
 of the IIB sigma model scalars.} complex 3-form  which includes the  R-R and the 3-form NS-NS 3-form field strengths, and $P$ is the twisted
1-form field strength of the dilaton and axion of the theory. Thus $H$ and $\xi$ are twisted complex 3- and 1-forms on the internal space $M^4$,
respectively.  We also introduce the frame  $(\bbe^+, \bbe^-, \bbe^z, \bbe^a, \bbe^i)$ as in (\ref{orthfr})
 and $\ell$ is the radius of $\mathrm{AdS}_6$. We  follow the conventions, notation and methodology of \cite{iibads} for investigating  IIB $AdS$ backgrounds.

To continue it is convenient to define
\bea
W_i={1\over6} \epsilon_i{}^{jkl} H_{jkl}~,
\eea
where $W$ is a complex (twisted) 1-form on $M^4$.
In such a case the Bianchi identities and the field equations that we shall use below  can be expressed as
\begin{eqnarray}
\nabla^i W_i=i Q_i W^i-\xi_i \bar W^i~,~~~
 d\xi =2 i Q \wedge \xi~,~~~ dQ = -i \xi \wedge \overline{\xi}~,
\end{eqnarray}
and
\begin{eqnarray}
\nabla_{[i} W_{j]}&=&-6 \partial_{[i}\log A W_{j]}+ i Q_{[i} W_{j]}+ \xi_{[i} \bar W_{j]}~,
 \cr
 {\nabla}^{i} \xi_{i} &=& -6 \partial^i \log A \, \xi_i + 2 i Q^{i} \xi_{i} - \frac{1}{4} W^2~,
\cr
 A^{-1} {\nabla}^2 A &=&  \frac{1}{8} \parallel W \parallel^2 - \frac{5}{\ell^2} A^{-2} - 5 (d\log A)^2~,
\cr
  {R}^{(4)}_{i j}
 &=&6 A^{-1} {\nabla}_{i} {\nabla}_{j} A+{3\over8} |W|^2 \delta_{ij}-{1\over2} W_{(i} \bar W_{j)}+2 \xi_{( i} \overline{\xi}_{j )}~,
 \label{ads6feq}
\end{eqnarray}
respectively, where $Q$ is the connection of the $U(1)$ bundle on the upper-half plane pulled back on $M^4$.

\subsection{The Killing spinors}

The KSEs of IIB supergravity can be solved along $AdS_6$ \cite{iibads} to yield an expression for the Killing spinor as
in  (\ref{kkk}),
where now $\sigma_\pm$ and $\tau_\pm$ are complex Weyl $Spin(9,1)$ spinors that depend only on the coordinates of $M^4$ and satisfy the projections
$\Gamma_\pm\sigma_\pm=\Gamma_\pm \tau_\pm=0$.

Next as in the IIA case define
\begin{eqnarray}
\Gamma_{(4)}= \Gamma_{zxyw}~,~~~\Gamma_{ijkl}=\epsilon_{ijkl}\Tilde\Gamma_{(4)}~,~~~
\end{eqnarray}
with the convention that
\bea
\Gamma_{(4)}\Tilde\Gamma_{(4)}\sigma_\pm =\pm \sigma_\pm~,
\label{g4g4}
\eea
and similarly for $\tau_\pm$, as these spinors. Using these conventions, the remaining independent KSEs are
\begin{eqnarray}
\nabla_i^{(\pm)}\sigma_\pm=0~,~~~\nabla_i^{(\pm)}\tau_\pm=0~,~~~{\cal A}^{(\pm)}\sigma_\pm=0~,~~~{\cal A}^{(\pm)}\tau_\pm=0~,
\label{gravdiliib}
\end{eqnarray}
and
\begin{eqnarray}
\Xi_\pm\sigma_\pm=0~,~~~\big(\Xi_\pm\pm {1\over\ell}\big)\tau_\pm=0~,
\label{xiiib}
\end{eqnarray}
where
\begin{eqnarray}
&&\nabla_i^{(\pm)}=  \nabla_i+ \Psi^{(\pm)}_i~,
\end{eqnarray}
with
\begin{eqnarray}
\Psi_i^{(\pm)}&=&\pm{1\over2} \partial_i\log A-{i\over2} Q_i- \big({1\over16} W_i\Tilde\Gamma_{(4)}+{3\over16} \Gamma^j{}_i W_j\Tilde\Gamma_{(4)}\big) C*~,
\cr
\Xi_\pm&=&\mp {1\over 2\ell}-{1\over2} \Gamma_z \partial_i A \Gamma^i+ {1\over 16} A \Gamma_z \slashed W\Tilde\Gamma_{(4)} C*~,~~
{\cal A}^{(\pm)}={1\over4} \slashed W\Tilde\Gamma_{(4)}+\slashed\xi C*~,
\end{eqnarray}
and $\nabla$ is the Levi-Civita connection on $M^4$.
The (\ref{gravdiliib}) KSEs are a suitable restriction of the gravitino and dilatino KSEs of IIB supergravity
on $\tau_\pm$ and $\sigma_\pm$ while the  (\ref{xiiib}) KSEs arise as integrability conditions of the solution of IIB KSEs along $AdS_6$.
Observe also that these KSEs imply that the $\sigma_+$ and $\tau_+$ Killing spinors are linearly independent.

The counting of Killing spinors is similar as that presented in more detail for $AdS_6$ solutions of 11-dimensional supergravity and so  IIB $AdS_6$ backgrounds preserve either 16 or 32 supersymmetries. It is known for sometime that there are no IIB $AdS_6$
backgrounds \cite{maxsusy}.  Therefore it remains  to explore the IIB $AdS_6$ backgrounds that preserve 16 supersymmetries.

\subsection{Non-existence theorems in IIB}

\subsubsection{The orthogonality of $\tau_+$ and $\sigma_+$ Killing spinors}

As in previous cases setting $\Lambda=\sigma_++\tau_+$ and upon using the gravitino KSE, one finds
\bea
\nabla_i \parallel \Lambda\parallel^2=-\partial_i \log A \parallel \Lambda\parallel^2+{1\over8} \mathrm {Re}\, \langle \Lambda, W_i\Tilde\Gamma_{(4)}  C*\Lambda\rangle~.
\eea
Next the algebraic  KSE  (\ref{xiiib}) gives
\bea
-\partial_i \log A \parallel \Lambda\parallel^2+{1\over8} \mathrm {Re}\, \langle \Lambda, W_i\Tilde\Gamma_{(4)}  C*\Lambda\rangle=2 \ell^{-1} A^{-1}
\mathrm {Re}\, \langle \tau_+, \Gamma_{iz} \sigma_+\rangle~.
\label{xyz}
\eea
Thus, one finds
\bea
\nabla_i \parallel \Lambda\parallel^2=2 \ell^{-1} A^{-1}
\mathrm {Re}\, \langle \tau_+, \Gamma_{iz} \sigma_+\rangle~.
\label{xxiib}
\eea
Furthermore observe that as a consequence of the gravitino KSE in (\ref{gravdiliib}), one has
\bea
\nabla^i(A\mathrm {Re}\, \langle \tau_+, \Gamma_{iz} \sigma_+\rangle)=0~.
\eea
Taking the divergence of (\ref{xxiib}) and using the above equation, one deduces that
\bea
\nabla^2\parallel \Lambda\parallel^2+ 2 A^{-1} \partial^i A \partial_i \parallel \Lambda\parallel^2=0~.
\label{maxiib1}
\eea
The maximum principle then gives that $\parallel \Lambda\parallel$ is constant and in turn
\bea
-\partial_i \log A \parallel \Lambda\parallel^2+{1\over8} \mathrm {Re}\, \langle \Lambda, W_i\Tilde\Gamma_{(4)}  C*\Lambda\rangle=0~,
\cr
\mathrm{Re}\, \langle \tau_+, \Gamma_{iz} \sigma_+\rangle=0~,
\label{mpcon}
\eea
where it is sufficient to assume the smoothness of  the fields and Killing spinors, and that $M^4$ is closed.
Furthermore the algebraic KSEs (\ref{xiiib}) give that
\bea
\mathrm{Re}\, \langle \tau_+, \sigma_+\rangle=0~.
\label{orthiib}
\eea
This establishes the orthogonality of $\tau_+$ and $\sigma_+$ spinors.

\subsubsection{A formula for the warp factor}

Observe that as a consequence of the first equation  in (\ref{mpcon}) as well as the orthogonality condition in (\ref{orthiib}) that
\bea
\mathrm{Re}\, \langle \sigma'_+,\Tilde\Gamma_{(4)}\sigma_+\rangle=0~,~~~\mathrm {Re}\, \langle \tau_+, W_i\Tilde\Gamma_{(4)}  C*\sigma_+\rangle=0~,
\label{xxx}
\eea
where we have used that  $\tilde\Gamma_{(4)}\sigma_+$ is a $\tau_+$ Killing spinor because of (\ref{g4g4}) and the relation between $\tau_+$ and $\sigma_+$ spinors.

Taking the derivative of the first equations in  (\ref{mpcon}), we obtain for $\Lambda=\sigma_+$ that
\bea
&&-\nabla^2 \log A  \parallel\sigma_+\parallel^2 - (\partial_i\log A)^2 \parallel\sigma_+\parallel^2
+
{1\over 64}  |W|^2  \parallel\sigma_+\parallel^2
\cr &&
~~~~+{3\over 64} \mathrm{Re}\, \langle\sigma_+, \bar W^i W^j \Gamma_{ij} \sigma_+\rangle
-{1\over8} \mathrm{Re}\, \langle\sigma_+, \xi_i \bar W^i\Tilde\Gamma_{(4)} C*\sigma_+\rangle=0~.
\eea
Moreover the dilatino KSE in (\ref{gravdiliib})  can be used to give
\bea
{1\over4} |W|^2 \parallel\sigma_+\parallel^2+{1\over 4} \mathrm{Re}\, \langle\sigma_+, \bar W^i W^j \Gamma_{ij} \sigma_+\rangle+
\mathrm{Re}\, \langle\sigma_+, \xi_i \bar W^i\Tilde\Gamma_{(4)} C*\sigma_+\rangle=0~.
\label{xxyy}
\eea
Eliminating the term that contains $\xi$ in the above two equations, we find
\bea
&&-\nabla^2 \log A  \parallel\sigma_+\parallel^2 - (\partial_i\log A)^2  \parallel\sigma_+\parallel^2
\cr &&
~~~~~~~~~~+
{3\over 64}  |W|^2 \parallel\sigma_+\parallel^2+{5\over 64} \mathrm{Re}\, \langle\sigma_+, \bar W^i W^j \Gamma_{ij} \sigma_+\rangle=0~.
\label{xxyyzz}
\eea
This is the key equation that we shall explore in what follows to establish non-existence theorems. Observe that apart from the last term,
it has the required form to apply the maximum principle on the warp factor $A$.

\subsubsection{A non-existence theorem for linearly dependent NSNS and RR fluxes}

IIB backgrounds have linearly dependent NSNS and RR fluxes if there exist nowhere vanishing
(twisted) complex function $f$  on $M^4$ such that
\bea
 \mathrm{Re}\big(f W_i\big)=0~.
\label{ldep}
\eea
This is the case for all IIB backgrounds that have either NSNS or RR 3-form fluxes and all their $SL(2, \bZ)$ duals.
For backgrounds for which the fluxes satisfy (\ref{ldep}), the last term in (\ref{xxyyzz}) vanishes. This follows easily after solving (\ref{ldep}) as
\bea
W_i=-{\bar f^2\over |f|^2} \bar W_i
\eea
and substituting into
\bea
\mathrm{Re}\, \langle\sigma_+, \bar W^i W^j \Gamma_{ij} \sigma_+\rangle=\mathrm{Re}\, \langle\sigma_+, \big(-{\bar f^2\over |f|^2}\big)\bar W^i  \bar W^j\Gamma_{ij} \sigma_+\rangle=0~.
\eea
As the last term in  (\ref{xxyyzz}) vanishes an application of the maximum principle reveals that the warp factor is constant and $W=0$.  This
in turn makes the field equation for the warp factor $A$ (\ref{ads6feq}) inconsistent and so such backgrounds do not exist.

Furthermore, it is required that smooth closed $AdS_6$ solutions must have active scalars.  Indeed if $\xi=0$, then the algebraic KSE in (\ref{gravdiliib}) implies
that
\bea
|W|^2 \parallel\sigma_+\parallel^2+\mathrm{Re}\, \langle\sigma_+, \bar W^i W^j \Gamma_{ij} \sigma_+\rangle=0~,
\eea
which can be used to eliminate the last term in (\ref{xxyyzz}). Then the maximum principle can apply to yield that $A$ is constant and  $W=0$ which in turn are inconsistent
with the  field equation for the warp factor $A$ (\ref{ads6feq}).

\subsubsection{A non-existence theorem for $\langle \sigma_+, C*\sigma_+\rangle\not=0$}

To demonstrate this non-existence theorem for smooth closed IIB $AdS_6$ solutions, we shall first show that either the bilinear $\langle \sigma_+, C*\sigma_+\rangle$  vanishes identically or it does not vanish anywhere on the internal space.    To begin, we evaluate the derivative of this bilinear and after using  the gravitino in (\ref{gravdiliib})  and  algebraic (\ref{xiiib})   KSEs, to  find that
\bea
\nabla_i\big(A^4 \langle \sigma_+, C*\sigma_+\rangle\big)=-i Q_i A^4 \langle \sigma_+, C*\sigma_+\rangle~.
\eea
This is a parallel transport equation which establishes the statement as $A$ does not vanish anywhere on the internal manifold.

Next
observe that the second equation in (\ref{xxx}) can be written as
\bea
\mathrm {Re}\,\langle \sigma_+, W_i   C*\sigma_+\rangle=0~.
\label{xxx2}
\eea
This condition  is as that in  (\ref{ldep}) with $f=\langle \sigma_+,    C*\sigma_+\rangle$. Therefore if $\langle \sigma_+, C*\sigma_+\rangle$ does not vanish identically, then the non-existence theorem follows from the results of the previous section. Observe
that  $\langle \sigma_+, C*\sigma_+\rangle$ satisfies the requirements  on the function $f$.

Incidentally observe that if the bi-linear $\alpha=\langle \sigma_+, \tilde\Gamma_{(4)} C*\sigma_+\rangle$ does not vanish anywhere on $M^4$, then there are no smooth closed solutions.
Indeed from (\ref{mpcon}), one finds that
\bea
 W_i ={16
\over |\alpha|^2} \partial_i \log A \parallel \sigma_+\parallel^2- {\bar\alpha^2
\over |\alpha|^2} \bar W_i~.
\label{wfgeom}
\eea
Substituting this into (\ref{xxyyzz}), the latter  can be rearranged such that the maximum principle applies  leading to $W=0$ which together with the field equations for the warp factor $A$ establishes the statement.

\subsubsection{A non-existence theorem for $\det\omega\not=0$}

So far we have seen that smooth closed IIB $AdS_6$ backgrounds exist provided that
\bea
\langle \sigma_+, C*\sigma_+\rangle=0~,
\label{cvan}
\eea
and $\alpha$ vanishes somewhere on the internal space.  However, there are additional restrictions on the spinor bilinears required for the existence of
such $AdS_6$ backgrounds. To find them, let us explore further the geometry of the internal space. A more systematic investigation of the spinor bilinears
can be found in appendices B and D.

To continue define
\bea
\omega_{ij}=\langle \sigma_+, \Gamma_{ij} \sigma_+\rangle~,~~~{}^*\omega_{ij}=\langle \sigma_+, \Gamma_{ij} \Tilde\Gamma_{(4)}\sigma_+\rangle~.
\eea
Observe that
$
\langle\sigma_+, \Gamma_{ij}\Tilde \Gamma_{(4)} \sigma_+\rangle=-{1\over2} \epsilon_{ij}{}^{kl} \langle\sigma_+, \Gamma_{kl} \sigma_+\rangle.
$
An application of the gravitino KSE reveals that
\bea
\nabla_k\omega_{ij}&=&- \partial_k \log A\langle\sigma_+, \Gamma_{ij} \sigma_+\rangle -{3i\over8}
[\mathrm{Im} \langle \sigma_+, W_i \delta_{kj}\Tilde\Gamma_{(4)} C* \sigma_+\rangle-(j,i)]~,
\cr
\nabla_k {}^*\omega_{ij}&=&- \partial_k \log A\langle\sigma_+, \Gamma_{ij} \Tilde\Gamma_{(4)}\sigma_+\rangle -{3i\over8}
\epsilon_{ijk}{}^m \,\mathrm{Im}\langle \sigma_+, W_m \Tilde\Gamma_{(4)} C* \sigma_+\rangle~.
\label{domega}
\eea
These in particular imply that
\bea
d(A\omega)=0~,~~~\nabla_i( A{}^*\omega_{jk})=\nabla_{[i}( A{}^*\omega_{jk]})~,
\label{clos}
\eea
ie $A{}^*\omega$ is a Killing-Yano tensor.

Furthermore, it is easy to show using the equation (\ref{domega}) that
\bea
\nabla_k (A^2 \omega_{ij} {}^*\omega^{ij})=0~.
\eea
Thus $A^2 \omega_{ij} {}^*\omega^{ij}=\mathrm{const}$.  If this constant does not vanish, then  $\omega$ has rank 4 everywhere on the internal manifold $M^4$.

On the other hand using (\ref{cvan}), one can show from the dilatino KSE that
\bea
W^i\,\, {}^*\omega_{ij}=0~.
\label{wgamma}
\eea
Thus if $A^2 \omega_{ij} {}^*\omega^{ij}\not=0$, and so both ${}^*\omega$ and $\omega$ are invertible, then $W$ will vanish everywhere on $M^4$.
Again in such a case there are no smooth closed $AdS_6$ solutions. This establishes our theorem.
Therefore if smooth closed $AdS_6$ solutions exist, they require that $\omega$ has rank 2 or less everywhere on the internal space.

\subsection{Symmetries and topology of  internal manifold}

Further restrictions on the internal space of smooth closed IIB $AdS_6$ backgrounds can be found by exploring the isometries of $AdS_6$ backgrounds.
We have established in appendix A that the internal space of $AdS_6$ backgrounds admits an $\mathfrak{so}(3)$ action generated by three\footnote{In fact there is the possibility
of a fourth for $Q=\Gamma_{123}$ but this vanishes identically because of (\ref{mpcon}).} vector fields
\bea
Z^Q_i= A \mathrm{Re}\,\langle \sigma_+, Q \Gamma_i \sigma_+\rangle~,
\label{kv}
\eea
for $Q=\Gamma_{z23}, -\Gamma_{z13}, \Gamma_{z12}$. It is then a consequence of
the KSEs that this action leaves all the fields invariant.

First let us demonstrate that this action is non-trivial.
For this we shall show that the $Z_Q$ cannot vanish identically for smooth closed solutions.  To see this first note that the Fierz identities in appendix B
give
\bea
\omega^2-2
|\alpha|^2 +2(\parallel \sigma_+ \parallel^2)^2&=&0~,
\cr
\sum_Q Z^2_Q
+2 A^2 |\alpha|^2
-2 A^2 \big(\parallel \sigma_+ \parallel^2\big)^2 &=&0~,
\label{nonvanx}
\eea
where $\alpha=\langle \sigma_+, {\tilde{\Gamma}}_{(4)} C* \sigma_+ \rangle$ as in previous sections and we have used that $\langle \sigma_+,  C* \sigma_+ \rangle=\langle \sigma_+, {\tilde{\Gamma}}_{(4)}  \sigma_+ \rangle=0$.

Now suppose that there are smooth closed solutions for which $Z_r$ vanish identically on the internal space.
Then as a consequence of the Fierz identities $\omega^2=0$ and so $\omega=0$. On the other hand  if $\omega$ vanishes, then an application of the maximum principle on  (\ref{xxyyzz}) implies
that there are no smooth closed $AdS_6$ solutions which is a contradiction. This establishes that for smooth closed solutions  $\mathfrak{so}(3)$ acts non-trivially on the internal space.

The groups with Lie algebra $\mathfrak{so}(3)$ act on 4-dimensional manifolds and the principal\footnote{These are the orbits with the smallest isotropy group.} orbits are either of co-dimension 1 or 2.  To continue, let us suppose that the principal orbits of $\mathfrak{so}(3)$ on $M^4$ have co-dimension 1.  As all fluxes are invariant up to a possible gauge $U(1)$ transformation and $Z^i_Q W_i=0$ (\ref{incon}), $W\wedge \bar W$  descends on the  space of orbits.  As the space of orbits
is 1-dimensional  $W\wedge \bar W$ vanishes.  Then we can apply the maximum principle on (\ref{xxyyzz}) to conclude that the warp factor is constant and $W=0$ which lead to an inconsistency
in the field equation for the warp factor.  Thus, {\it  there are no smooth closed $AdS_6$ backgrounds  with  principal $\mathfrak{so}(3)$ orbits on the internal space of  co-dimension 1.}

It remains to investigate the case that for which the principal orbits of $\mathfrak{so}(3)$ action on $M^4$ have co-dimension 2. For this let us assume
that the $\mathfrak{so}(3)$ action on $M^4$ can be integrated to an effective  $SU(2)$ or  $SO(3)$ on $M^4$, and that $M^4$ is oriented. Oriented 4-dimensional manifolds admitting an effective non-abelian group action have been classified in \cite{parker} and references within, see also \cite{hillman}.  The principal orbits are 2-spheres. These can degenerate to either points or to the   2-dimensional real projective
space $\bR P^2$. The space of orbits is a 2-dimensional surface $F$ with boundary which is a union of circles. The principal orbits degenerate at the boundary circles.
Spin oriented 4-dimensional manifolds $M^4$ which admit an effective  $SU(2)$ or  $SO(3)$ with principal co-dimension 2 orbits include\footnote{There are some other possibilities, see appendix C.}
those presented in (\ref{diffma}).
 For $M(F)$, the surface $F$ has no boundary and
the principal orbits do not degenerate. If $F$ is oriented, then $M(F)=F\times S^2$. However we shall demonstrate later that there are no smooth closed $AdS_6$ backgrounds
with internal space  $M^4$ that has topology $F\times S^2$.

\subsection{Aspects of geometry}

To explore the geometry of the internal space of IIB $AdS_6$ backgrounds, let us assume that $\langle \sigma_+,  C* \sigma_+ \rangle=\langle \sigma_+, {\tilde{\Gamma}}_{(4)}  \sigma_+ \rangle=0$
and  $\det\omega=0$.  Observe that (\ref{wgamma}) can be rewritten as
\bea
\label{ortho1}
W\wedge \omega=0~,
\eea
and the algebraic and dilatino KSEs give
\bea
\xi\wedge \omega=dA\wedge \omega=0~.
\eea
The 2-form $A\omega$  descends on the space of orbits $F$ as ${\cal L}_{Z^Q}A\omega=0$ and $i_{Z^Q} \omega=0$.  The latter follows from (\ref{fierz3}).

Note that $\omega$  gives rise to a Hermitian structure on $F$ away from the fixed points.  This can be seen from the Fierz identity (\ref{fierz11}) which can be re-written as
\bea
\label{fierz11x}
A^2 \omega_{ik}  \omega_j{}^k
&=& \sum_Q Z_i^Q
 Z^Q_j
+
A^2 \bigg(|\alpha|^2
- \big(\parallel \sigma_+ \parallel^2\big)^2 \bigg) \delta_{ij}~.
\eea
If there are points for which $|\alpha|=\parallel \sigma_+ \parallel^2$, $\omega$ is not a Hermitian form on $F$.  Incidentally, the points for which
 $|\alpha|=\parallel \sigma_+ \parallel^2$ are the fixed points of the group action as can be seen from (\ref{fierz7}).

In addition, setting as before $\alpha=\langle \sigma_+, \tilde\Gamma_{(4)} C*\sigma_+\rangle$, one finds that
\bea
\nabla_i\alpha&=& -\partial_i\log A \alpha-iQ_i \alpha
+{1\over8} \bar W_i \parallel\sigma_+\parallel^2+{3\over8} \bar W_m \omega^m{}_i~.
\eea
Using the algebraic KSE $\Xi$ to eliminate the last term, we find that
\bea
\nabla_i( A^4\alpha)=-iQ_i  A^4 \alpha
+{1\over2} A^4 \bar W_i \parallel\sigma_+\parallel^2-{3\over\ell} A^{3} \langle \sigma_+,\Gamma_z \Gamma_i  \tilde\Gamma_{(4)} C*\sigma_+\rangle~.
\eea
This condition can be solved for $W$ to  determine the flux in terms of the geometry of the internal space.

\subsection{A non-existence theorem for $M^4=F\times S^2$}

We shall now demonstrate that there are no smooth closed $AdS_6$ solutions for which the internal space has topology\footnote{It is allowed though for the metric
  on $M^4$ to be a warped product of the metric of $S^2$ with that of $F$.} $M^4=F\times S^2$, where $F$ is an  oriented surface.
To see this, observe that in this case  the action of $\mathfrak{so}(3)$ has no fixed points. The orbits are 2-spheres and all are principal.  As a result, there are no points in $M^4$ for which all Killing vector fields
$Z^Q$ vanish simultaneously.  Then the second equation in (\ref{nonvanx}) implies that
\bea
|\alpha|-\parallel\sigma_+\parallel^2<0~.
\eea
In turn the first equation in  (\ref{nonvanx}) implies that $\omega$ is nowhere vanishing.  As $A$ is nowhere vanishing, the 2-form $A^n \omega$ is nowhere vanishing
as well for $n\in \bZ$.  Furthermore $A^n \omega$ descends on $F$ and as $A^n \omega$ is nowhere vanishing, one has
\bea
\int_F \, A^n \omega\not=0~.
\label{contra}
\eea
On the other hand, an application of the gravitino KSE (\ref{gravdiliib}) reveals that
\bea
&&\nabla_i \langle \sigma_+, \Gamma_{aj} \sigma_+ \rangle=-\partial_i\log A \langle \sigma_+, \Gamma_{aj} \sigma_+ \rangle+{i\over 8} \mathrm{Im} \langle \sigma_+, W_i \Gamma_{aj}\tilde\Gamma_{(4)}C* \sigma_+ \rangle
\cr &&~~~~~~
+{3i\over8} \mathrm{Im}\langle \sigma_+, W_j \Gamma_{ai}\tilde\Gamma_{(4)}C* \sigma_+ \rangle-{3i\over8} \delta_{ij} \mathrm{Im}\langle \sigma_+, \Gamma_{a} W_k \Gamma^k\tilde\Gamma_{(4)}C* \sigma_+ \rangle~,
\eea
for $a=1,2,3,z$.
Using this and the KSE (\ref{xiiib}) we obtain
\bea
 A^2 \omega={\ell\over2} d (A^3 \zeta)~,
\eea
where $\zeta_i=\langle \sigma_+, \Gamma_{zi} \sigma_+ \rangle$.  Pulling this equation back on $F$ with a global section of the trivial fibration $F\times S^2$, we find upon
application of Stoke's theorem that
\bea
\int_F  A^2 \omega=0~.
\eea
This is in contradiction to (\ref{contra}) for $n=2$.  We therefore conclude that there are no  smooth closed $AdS_6$ solutions with internal manifolds that have topology
$F\times S^2$.

\subsection{Conditions for the existence of $AdS_6$  IIB solutions   }

We have seen that the existence of smooth closed IIB $AdS_6$ backgrounds is rather restricted.  However, we have not been
able to establish as strong a result as that for $D=11$ and (massive) IIA backgrounds.  If smooth closed IIB $AdS_6$ backgrounds exist, the statements of non-existence theorems that we have established  can turn to necessary conditions that  such backgrounds must satisfy. Of course the issue of existence of smooth closed IIB $AdS_6$ backgrounds can be settled with an example.  However to our knowledge
all solutions that have been found in the literature \cite{andy}-\cite{dhoker2} so far are singular. In particular those found in \cite{dhoker2} are singular because the warp
 factor $A$ of the $AdS_6$ subspace vanishes at some points on the internal space.  As it has been shown in \cite{iibads} that  irrespective of the frame chosen,  $A$ is related to the length of a Killing spinor \cite{iibads}, $\parallel \sigma_-\parallel^2= c^2 A^2$, where $c$ is a constant. For smooth closed $AdS_6$ backgrounds, the Killing spinors must be nowhere vanishing, and so $A\not=0$ everywhere on the internal space. Here we shall summarize the necessary conditions for existence
of smooth closed IIB $AdS_6$ backgrounds.

We have seen that the bilinears of  Killing spinor $\sigma_+$ must satisfy the following conditions
\bea
\langle \sigma_+, C*\sigma_+\rangle=0~,~~~~\det \omega=0~,~~~\langle \sigma_+, \tilde\Gamma_{(4)}\sigma_+\rangle=0~.
\eea
Violation of either of the first two conditions leads to a non-existence result while the third condition is a consequence of
the orthogonality of $\sigma_+$ and $\tau_+$ Killing spinors.

Furthermore the bilinear
\bea
\langle \sigma_+, \tilde\Gamma_{(4)} C*\sigma_+\rangle~,
\eea
must vanish at some points of the internal space $M^4$.  Otherwise again a non-existence theorem for smooth closed $AdS_6$
backgrounds can be established.

Another necessary condition for the existence of smooth closed IIB $AdS_6$ backgrounds is that they must have active scalars.  If
the IIB scalars are constant, then one can demonstrate that there is a non-existence theorem for such solutions.

In addition, it has been shown in appendix D that for smooth closed  $AdS_6$ backgrounds, the following spinor bilinears also vanish
\bea
\langle \sigma_+, \Gamma_{12}\sigma_+\rangle=\langle \sigma_+, \Gamma_{13}\sigma_+\rangle=\langle \sigma_+, \Gamma_{23}\sigma_+\rangle=0~,
\eea
otherwise such backgrounds cannot exist.
This is equivalent to the statement that all $\sigma$-type  Killing spinors are orthogonal over the complex numbers.  This is despite
the fact that the KSEs of IIB supergravity are linear over the real numbers.

Moreover we have demonstrated that if smooth closed IIB $AdS_6$ solutions exist, the internal space has principal orbits of $\mathfrak{so}(3)$ of
co-dimension 2. Otherwise again a non-existence theorem can be established. Combining this with the classification results of \cite{parker},
it is possible to determine the diffeomeorphic type of the internal space that includes manifolds as those in (\ref{diffma}).  Using these
classification results, we also excluded the existence of smooth closed IIB $AdS_6$ solutions with internal space which has topology $F\times S^2$,
where $F$ is an oriented surface. This restricts further the possible topologies
of smooth  closed IIB $AdS_6$ backgrounds to $S^4$, $~p (S^1\times S^3)\# q (S^1\times \bR P^3)$ and  $M(F)$, where $F$ is a non-oriented surface.

Amongst these, an interesting topology for an internal space is $~p (S^1\times S^3)\# q (S^1\times \bR P^3)$.  The space of orbits is a surface $F$ with boundary
$\partial F$ which is a union of circles. If $n$ is the number of circles that $S^2$ degenerates to a point, then $q=\mathrm {rank} H_0(\partial F)-n$ and
$p=\mathrm {rank} H_1( F)-q$. For $q=0$ and $p=1$, $F$ is a cylinder as expected and $n=2$, while for $q=1$ and $p=0$ again $F$ is a cylinder but
now $n=1$. The internal spaces of  examples explored in \cite{dhoker, dhoker2} may have such topology.

\vskip 0.5cm
\noindent{\bf Acknowledgements} \vskip 0.1cm
\noindent   We would like to thank  Michael Gutperle, Neil Lambert,  Alessandro Tomasiello and Kostas Sfetsos  for helpful correspondence.  JG is supported by the STFC grant, ST/1004874/1. GP is partially supported by the  STFC rolling grant ST/J002798/1.
\vskip 0.5cm

\setcounter{section}{0}
\setcounter{subsection}{0}

\appendix{Symmetries}

\subsection{Killing Vector fields}

Let us define the 1-form spinor bilinears as in (\ref{kv}). Then using the gravitino KSE, one finds
\bea
\nabla_i Z^Q_j=-{3\over8} A \epsilon_{ij}{}^{kl} \mathrm{Re}\,\langle \Lambda, W_k Q \Gamma_l C*\Lambda\rangle~,
\label{dzeta}
\eea
where $Q=\Gamma_{z23}, -\Gamma_{z13}, \Gamma_{z12}$.  Potentially, there is an additional  vector bilinear for
$Q=\Gamma_{123}$ but it vanishes identically because of (\ref{mpcon}). However note that to derive (\ref{mpcon}) one uses
the maximum principle and so the vanishing of this  vector field requires a global condition.
It follows directly that the associated vector fields are Killing. Furthermore, one finds from
 the algebraic KSEs that
\bea
Z_k^Q W^k=0~,~~~Z_k^Q \bar \xi^k=0~,~~~Z^Q_k \partial^kA=0~.
\label{incon}
\eea
Next after using the Einstein equation, one can show that
\bea
\nabla^2 |Z^Q|^2={1\over2}|dZ^Q|^2-6 \partial^k\log A \partial_k |Z^Q|^2-{3\over4} |W|^2 |Z^Q|^2~.
\eea
Notice however that the right-hand-side of this equation is not definite to apply the maximum principle.

The algebra of the three  vector fields is
\bea
[Z^{Q_1}, Z^{Q_2}]=-{3\over \ell} \parallel\sigma_+\parallel^2\, Z^{Q_3}~,
\eea

where $Q_1=\Gamma_{z23}, Q_2=-\Gamma_{z13}, Q_3=\Gamma_{z12}$ and cyclic in the $Q$'s for the other commutators.  Therefore it is isomorphic to $\mathfrak{so}(3)$.
To prove this one uses (\ref{dzeta}), (\ref{incon}), the algebraic KSE $\Xi$ in (\ref{xiiib}) on $\sigma_+$ as well as (\ref{fierz7}). This is significant
as 4-dimensional manifolds admitting effective non-abelian group actions have been classified.

\subsection{Notation and conventions}

Our form conventions are as follows. Let $\omega$ be a k-form, then
\bea
\omega={1\over k!} \omega_{i_1\dots i_k} dx^{i_1}\wedge\dots \wedge dx^{i_k}~,~~~\omega^2_{ij}= \omega_{i\ell_1\dots \ell_{k-1}} \omega_{j}{}^{\ell_1\dots \ell_{k-1}}~,~~~
\omega^2= \omega_{i_1\dots i_k} \omega^{i_1\dots i_k}~.
\eea
We also define
\bea
{\slashed\omega}=\omega_{i_1\dots i_k} \Gamma^{i_1\dots i_k}~, ~~{\slashed\omega}_{i_1}= \omega_{i_1 i_2 \dots i_k} \Gamma^{i_2\dots i_k}~,~~~\slashed{\gom}_{i_1}= \Gamma_{i_1}{}^{i_2\dots i_{k+1}} \omega_{i_2\dots i_{k+1}}~,
\eea
where the $\Gamma_i$ are the Dirac gamma matrices.

The inner product $\langle\cdot, \cdot\rangle$ we use on the space of spinors is that for which space-like gamma matrices are Hermitian while time-like gamma
matrices are anti-hermitian, ie the Dirac spin-invariant inner product is $\langle\Gamma_0\cdot, \cdot\rangle$.  For more details on our  conventions
see \cite{mads, iiaads, iibads}.

\section{IIB Fierz Identities}

\setcounter{equation}{0}

In the investigation of the geometry of the internal spaces of IIB  $AdS_6$ backgrounds, we have used some Fierz identities.
These are listed below without assuming the vanishing of any of the spinor bilinears which are necessary for the existence of smooth closed IIB $AdS_6$ backgrounds.
So the identities below are valid in general. However, we shall comment on their consequences in the special cases of smooth closed IIB $AdS_6$ backgrounds.

The first Fierz identity gives $\omega^2$ in terms of scalar bilinears as
\bea
\label{fierz1}
\langle \sigma_+, \Gamma_{ij} \sigma_+ \rangle
\langle \sigma_+, \Gamma^{ij} \sigma_+ \rangle
&=&2|\langle \sigma_+, {\tilde{\Gamma}}_{(4)} C* \sigma_+ \rangle|^2 -2(\parallel \sigma_+ \parallel^2)^2
\nonumber \\
 &-&2|\langle \sigma_+, {\tilde{\Gamma}}_{(4)} \sigma_+ \rangle|^2
+2 |\langle \sigma_+, C* \sigma_+ \rangle|^2~.
\eea
Observe that the length of $\omega$ can vanish for  smooth closed IIB $AdS_6$ backgrounds whenever ${\tilde{\Gamma}}_{(4)} C* \sigma_+=\sigma_+$ as a consequence
of the Cauchy-Schwarz inequality.

Next, one finds that
\bea
\label{fierz2}
\epsilon^{ijmn} \langle \sigma_+, \Gamma_{ij} \sigma_+ \rangle \langle \sigma_+, \Gamma_{mn} \sigma_+ \rangle
&=&-8 {\rm Re} \bigg( \langle C* \sigma_+, \sigma_+ \rangle \langle
\sigma_+, {\tilde{\Gamma}}_{(4)} C* \sigma_+ \rangle\bigg)
\nonumber \\
&+&
8 \parallel \sigma_+ \parallel^2 \langle \sigma_+, {\tilde{\Gamma}}_{(4)} \sigma_+ \rangle~.
\eea
For smooth closed IIB $AdS_6$ solutions, this implies that $\omega$ must have rank strictly less than 4.  This was derived already using
a different reasoning.

The Fierz identity
\bea
\label{fierz3}
\langle \sigma_+, \Gamma_{ij} \sigma_+ \rangle
\langle \sigma_+, Q \Gamma^j \sigma_+ \rangle
&=& {1 \over 2} \langle C* \sigma_+, \sigma_+ \rangle
\langle \sigma_+, Q \Gamma_i C* \sigma_+ \rangle
\nonumber \\
&-&{1 \over 2} \langle \sigma_+, C* \sigma_+ \rangle
\langle Q \Gamma_i C* \sigma_+, \sigma_+ \rangle
\nonumber \\
&+& \langle \sigma_+, {\tilde{\Gamma}}_{(4)} \sigma_+
\rangle \langle \sigma_+, {\tilde{\Gamma}}_{(4)} Q \Gamma_i \sigma_+ \rangle~,
\eea
is instrumental to show that $A\, \omega$ descends on the space of orbits of $\mathfrak{so}(3)$ action on the internal space $M^4$ for smooth closed backgrounds.

The Fierz identity
\bea
\label{fierz7}
\sum_{a=1}^4 \langle \sigma_+, Q_a \Gamma_i \sigma_+ \rangle
\langle \sigma_+, Q_a \Gamma^i \sigma_+ \rangle
+2 | \langle \sigma_+, {\tilde{\Gamma}}_{(4)} C* \sigma_+ \rangle|^2
-2 \big(\parallel \sigma_+ \parallel^2\big)^2
\nonumber \\
+2 \langle \sigma_+, {\tilde{\Gamma}}_{(4)} \sigma_+ \rangle^2
-2 |\langle \sigma_+, C* \sigma_+ \rangle |^2 =0~,
\eea
has been instrumental in showing that the $\mathfrak{so}(3)$  acts non-trivially on the internal space of smooth closed $AdS_6$ backgrounds.

The Fierz identity
\bea
\label{fierz11}
\langle \sigma_+, \Gamma_{ik} \sigma_+ \rangle \langle \sigma_+, \Gamma_j{}^k \sigma_+ \rangle
&=& \sum_{a=1}^4 \langle \sigma_+, Q_a \Gamma_i \sigma_+ \rangle
\langle \sigma_+, Q_a \Gamma_j \sigma_+ \rangle
\nonumber \\
&+&
\bigg(| \langle \sigma_+, {\tilde{\Gamma}}_{(4)} C* \sigma_+ \rangle|^2
- \big(\parallel \sigma_+ \parallel^2\big)^2 \bigg) \delta_{ij}
\eea
has been used to investigate the existence of a Hermitian structure on the space of orbits of the $\mathfrak{so}(3)$ action on the internal space $M^4$.

\section{4-Manifolds with Large Symmetry Groups}

It has been known for sometime \cite{parker} that closed oriented 4-dimensional manifolds that support  an effective action of $SO(3)$ or $SU(2)$ are diffeomorphic to

\begin{enumerate}

\item $S^4$ or $\pm \bC P^2$,

\item connected sums  $p (S^1\times S^3)\# q (S^1\times \bR P^3)$,

\item bundles over $S^1$ with fibre $SU(2)/D$, where $D$ is a discrete subgroup,

\item bundles over  surfaces with fibre $S^2$, and

\item certain quotients of $S^2$-bundles over surfaces with involutions.

\end{enumerate}

From these those collected in (\ref{diffma}) are spin, closed, and  $SO(3)$ or $SU(2)$ have co-dimension 2 principal orbits. The  $SO(3)$ or $SU(2)$ group action on $M(F)$ have orbits 2-spheres  $S^2$ and are all principal. $M(F)$ is uniquely characterized as the 2-sphere bundle over a surface $F$ that admits a spin structure. $M(F)=F\times S^2$, if $F$ is oriented. The spaces $p (S^1\times S^3)\# q (S^1\times \bR P^3)$ arise in the case that the principal orbit $S^2$ degenerates to several other
special orbits. For oriented 4-manifolds $S^2$ either  collapses to a point or  degenerates to  $\bR P^2$. In both cases, the special orbits are not isolated but rather lie on a circle. In particular in $S^1\times S^3$ the $S^2$ fibre collapses to a  point at   two different circles, while in $S^1\times \bR P^3$ the $S^2$ fibre collapses to point at a circle and  becomes $\bR P^2$ at another.  There are some more oriented 4-manifolds admitting co-dimension 2 principal orbits.  An example is $P(F)$ for $F$ a 2-dimensional surface for which at each boundary circle the principal $S^2$ orbit
degenerates to $\bR P^2$. If $F$ is a two 2-disk $B^2$, then $P(F)=M(\bR P^2)$. For other 2-dimensional surfaces $F$ with boundaries, $P(F)$ gives another class of oriented 4-dimensional manifolds
 which exhibit co-dimension 2 principal orbits. However, we have not been able to establish whether $P(F)$ are spin for $F\not=B^2$.

Note that in IIB, the Killing spinors have a Spin$_c$ structure rather than a Spin structure.  In particular the spinors are twisted with an additional $U(1)$ bundle which is the pull back on the spacetime of  canonical $U(1)$ bundle that arises after viewing  the target space of  IIB  scalars   as a  $SU(1,1)/U(1)$ coset space. As this coset space is the upper half plane,  it is contractible unless one considers further identifications which we shall not investigate here. As a result the $U(1)$ bundle is topologically trivial and so this Spin$_c$ structure is equivalent to a standard Spin structure.

\section{Further conditions on scalar spinor bi-linears}

\setcounter{equation}{0}

In this appendix, we shall further investigate some properties of the scalar spinor bilinears, and
establish more vanishing conditions for some of these. The  scalar bilinears are
\bea
\langle \sigma_+,  \sigma_+ \rangle~,~~~\langle \sigma_+, \tilde\Gamma_{(4)} \sigma_+ \rangle~,~~~\langle \sigma_+, \Gamma_{ab} \sigma_+ \rangle~,~~~\langle \sigma_+,  C*\sigma_+ \rangle~,~~~
\langle \sigma_+,  \tilde\Gamma_{(4)}C*\sigma_+ \rangle~,
\eea
where $a,b=1,2,3,z$. The last two bilinears are twisted with a $U(1)$.
We have already shown that for smooth closed $AdS_6$ backgrounds  $\parallel \sigma_+\parallel$ is constant and  $\langle \sigma_+, \tilde\Gamma_{(4)} \sigma_+ \rangle=\langle \sigma_+,  C*\sigma_+ \rangle=0$.  We shall assume that  these three conditions are valid throughout this appendix. So it remains to investigate the bilinears $\langle \sigma_+, \Gamma_{ab} \sigma_+ \rangle$
and $\langle \sigma_+,  \tilde\Gamma_{(4)}C*\sigma_+ \rangle$.

For the latter scalar bilinear, an application of the gravitino KSE (\ref{gravdiliib}) reveals that
\bea
\nabla_i \langle \sigma_+,  \tilde\Gamma_{(4)}C*\sigma_+ \rangle&=&-\partial_i\log A \langle \sigma_+,  \tilde\Gamma_{(4)}C*\sigma_+ \rangle-iQ_i \langle \sigma_+,  \tilde\Gamma_{(4)}C*\sigma_+ \rangle
\cr &&
+{1\over8} \bar W_i \parallel \sigma_+\parallel^2+{3\over8} \bar W_m \langle \sigma_+,  \Gamma^m{}_i\sigma_+ \rangle~,
\eea
which may be  used to determine the flux $W$ in terms of geometry.

For the $\langle \sigma_+, \Gamma_{ab} \sigma_+ \rangle$ scalar bilinear,
we have after an application of the gravitino KSE  (\ref{gravdiliib}) that
\bea
\nabla_i\langle \sigma_+, \Gamma_{ab} \sigma_+ \rangle=-\partial_i \log A \langle \sigma_+, \Gamma_{ab} \sigma_+ \rangle+ {3i\over8} \mathrm{Im} \langle \sigma_+, W_j \Gamma_{ab} \Gamma^j{}_i \tilde\Gamma_{(4)} C* \sigma_+ \rangle
\eea
Setting $a=r, b=s$  for $r,s=1,2,3$ and after using the KSE (\ref{xiiib}), we find the   identities
\bea
&&{A^{-1}\over\ell}  \langle \sigma_+, \Gamma_{rszi} \sigma_+ \rangle+\partial_j\log A \langle \sigma_+, \Gamma_{rs} \Gamma^j{}_i \sigma_+ \rangle-{1\over8}\mathrm{Re} \langle \sigma_+, \Gamma_{rs} \Gamma^j{}_i W_j\tilde\Gamma_{(4)} C*\sigma_+ \rangle=0~,
\cr
&&\partial_i\log A\, \mathrm{Im}  \langle \sigma_+, \Gamma_{rs}  \sigma_+\rangle +{1\over8}\mathrm{Im} \langle \sigma_+, \Gamma_{rs} \Gamma^j{}_i W_j\tilde\Gamma_{(4)} C*\sigma_+ \rangle=0~,
\label{scal3}
\eea
and
\bea
\partial_i (A^4 \langle \sigma_+, \Gamma_{rs} \sigma_+ \rangle)=0~.
\label{scal4}
\eea
We shall now demonstrate that
\bea
\langle \sigma_+, \Gamma_{rs}\sigma_+\rangle=0~,
\label{vanconxx}
\eea
for smooth closed $AdS_6$ backgrounds.  To prove this,
take the derivative of (\ref{scal3}) and use the gravitino KSE (\ref{gravdiliib})  and field equation for $W$, to obtain an expression for $\nabla^2\log A$ which depends on the scalars $\xi$.
Eliminating the $\xi$ terms using the algebraic KSE in (\ref{gravdiliib}), we find an expression for  $\nabla^2\log A$ which depends both on $|W|^2$ and $W\wedge \bar W$. It turns out that
the last term can be eliminated using the the algebraic KSE for $\Xi$ in (\ref{xiiib}).   The resulting expression is
\bea
&&\big [\nabla^2 \log A+6 (\partial\log A)^2 -{1\over8} |W|^2 \big ]\langle \sigma_+, \Gamma_{rs}\sigma_+\rangle
=0~.
\eea

Observe that because of (\ref{scal4}), $\mathrm{Im}\langle \sigma_+, \Gamma_{rs}\sigma_+\rangle$ has a definite sign. If $\langle \sigma_+, \Gamma_{rs}\sigma_+\rangle\not=0$,
an application of the maximum principle reveals that $A$ is constant and $W=0$.  There are no such smooth closed AdS$_6$ solutions. This establishes that (\ref{vanconxx})
must be satisfied on smooth closed $AdS_6$ backgrounds.

\setcounter{equation}{0}

\vskip 0.5cm
\noindent{\bf Acknowledgements} \vskip 0.1cm
\noindent   JG is partially supported by the STFC Consolidated Grant ST/L000490/1. GP is partially supported by the  STFC rolling grant ST/J002798/1.
\vskip 0.5cm

%

\end{document}